# Shift of nanodroplet and nanocluster size distributions induced by dopant pick-up statistics

Marie Riddle,[1] Angel I. Pena Dominguez,[2] Benjamin S. Kamerin,[2] Vitaly V. Kresin[2]

[1]*Department of Physics, Princeton University, Princeton, NJ 08544-0708, USA*

[2]*Department of Physics and Astronomy, University of Southern California,*
*Los Angeles, California 90089-0484, USA*

**Abstract**

In pick-up experiments using nanodroplet and nanocluster beams, the size distribution of hosts carrying a specified number of dopants changes when the vapor density in the pick-up region is altered. This change, analyzed here, has quantitative consequences for the interpretation of data that are sensitive to host size, such as mass spectrometric, spectroscopic, and deflection measurements.

The pick-up method is common in cluster science.[1] By passing a beam of nanoclusters or nanodroplets through a region filled with a dilute atomic or molecular vapor, the particles' exterior or interior can be populated with one or more dopants. Helium nanodroplets[2] and rare-gas and water nanoclusters[3,4] are common platforms for this technique. More recently, hydrated-acid[4,5] and polyaromatic hydrocarbon[6] nanoclusters and nanoparticles also have been used as hosts.

When successive pick-up events are independent, which is typically the case, the probability for a cluster of size $N$ to pick up $k$ dopants is given by Poisson statistics:

$$\mathcal{P}(k\,|\,N) = \frac{\lambda^k e^{-\lambda}}{k!}. \qquad (1)$$

Here $\lambda = nl\sigma_N$, where $n$ is the number density of molecules in the vapor, $l$ is the path length through the pick-up region, and $\sigma_N$ is the pick-up collision cross section. The latter is frequently approximated by the cross-sectional area of the cluster, so that[7] $\lambda \propto N^{2/3}$. For conciseness, we will refer to both nanoclusters and nanodroplets as "clusters."

Crucially, a cluster beam is not monodisperse. It contains a distribution of cluster sizes, $P(N)$. The distribution is characterized by a mean $\bar{N}$ and a standard deviation $s$. (In many cases this distribution is taken to be log-normal.[2,10])

It is inviting to apply Eq. (1) to cluster mass spectrometry or spectroscopy by treating $\lambda$ as a fixed quantity representing the average cluster size produced in the source: $\lambda \to \bar{\lambda} = nl\bar{\sigma}$. If that were accurate, the number $k$ of dopants per cluster could be reliably deduced from the variation of the intensity of a dopant peak in the mass spectrum with pick-up vapor pressure. Or conversely, the average cluster size could be determined by fitting the intensity of a $k$-mer impurity signal to the Poisson curve.

However, this turns out to be an unreliable approximation. For example, ion intensity curves do not necessarily fall onto a single Poissonian, [11,12] while fits to cluster sizes appear to vary depending on the dopant[13] or are inconsistent with other data.[14] It is the aim of this note to point out a simple but seemingly little discussed effect which can significantly impact measurements when the signal from one selected dopant is monitored as a function of pick-up vapor pressure.

Note that this evaluation is complementary to well-known simulations of pick-up statistics[2,15-17] in that their focus is usually on the distribution of dopants transported by clusters of a given size, whereas here we are interested in the distribution of cluster sizes that are transporting a given fixed number of dopants.

Suppose, for example, that the measurement monitors the signal intensity from a dopant monomer. When the vapor density is very low, the majority of clusters will remain undoped, hence the fraction that succeeds in picking up a dopant will tend to have cross sections that are larger than the beam average. Conversely, when vapor becomes dense, the average clusters will be carrying $k>1$ dopants, hence the singly-doped ones will tend to be smaller than the average. This, in turn, can affect the strength of the experimental signal being monitored.

An overt example of such behavior was noticed during our experiments[18-21] on electric and magnetic deflection of helium nanodroplets doped with polar and magnetic molecules. While detecting the deflection of singly-doped droplets, we observed the initially puzzling pattern that this deflection increased as the vapor pressure in the pick-up cell was being raised. An example is shown in Appendix 1. It emerged that the explanation was the one described above: the subset of clusters carrying only one dopant was becoming smaller on average. As such, this subset was experiencing a greater sideways acceleration from the deflecting field.

It is possible to determine the probability distribution $P(N|k)$ of cluster sizes $N$ specified to carry a given number of dopants, $k$. One way is by use of Bayes' theorem, which in the present case reads as follows:

$$P(N|k) = \frac{P(N)\mathcal{P}(k|N)}{P(k)} \tag{2}$$

The functions in the numerator are the ones defined above. The function in the denominator is the *a priori* probability that $k$ dopants were picked up. It is given by the convolution of $P(N)$ and Eq. (1). Thus, for example, the average size of clusters specified to carry $k$ dopants becomes

$$\langle N \rangle_k = \int N P(N|k) dN = \frac{\int N P(N) \mathcal{P}(k|N) dN}{\int P(N) \mathcal{P}(k|N) dN} \tag{3}$$

If the starting cluster size distribution and the pick-up region parameters are known, this expression defines the conditional distribution and its mean value. The key point of this note is that

$$\langle N \rangle_k \neq \bar{N}. \tag{4}$$

As an example, Fig. 1 shows the difference between these quantities for an experiment detecting single dopants ($k=1$) carried by a helium nanodroplet beam. The pick-up cross section is $\sigma_N = \pi R^2$ with[22] $R = 2.2 N^{1/3}$ Å. The original nanodroplet size distribution is taken to be log-normal with $\bar{N} = 10^4$ and $s = 9 \times 10^3$ (see Appendix 2 for a brief recap of the functional form of this distribution). The difference is plotted as a function of the nominal mean number of nanodroplet collisions in the pick-up region, $\bar{k}$. (For a Poisson distribution, this is the same as the parameter $\bar{\lambda}$ defined above.) The figure plots $\langle N \rangle_1$ calculated both from Eq. (3) and via a direct simulation of the pick-up process. The two are in perfect agreement, and the deviation of the corresponding population from the original mean size $\bar{N}$ is precisely in accord with the above reasoning.

(Note that the plot and the discussion does not include the effect of nanodroplet shrinkage due to post-collision evaporation, which can be incorporated in a straightforward manner and in essence only shifts the curve slightly downward.)

Analogously, Fig. 2 illustrates the difference in nanodroplet size distributions. Here the original distribution is compared with the populations carrying $k=1$ dopant after traversing the pick-up region at low or high vapor densities. The shift between the original and the conditional distributions is evident.

Therefore for experiments which focus on the signal intensity of specific dopant size(s) while the number of pick-up collisions is varied, it is important to be mindful of the fact that concurrently a change takes place in the ensemble of clusters carrying this dopant. This concurrent shift will affect measurements in cases when the detected signal depends on the size of the host cluster. Examples include not only beam deflection measurements but also ion yields in electron-impact ionization or high-intensity laser excitation[2,23] where the ionization, charge exchange, and excitation efficiencies are sensitive to the size of the host. In these cases the assumption of an unchanging carrier nanodroplet or nanocluster size $\bar{N}$ will lead to inaccuracies. The deviations of dopant ion intensities from a simple Poisson dependence, Eq. (1), which were mentioned in the introduction, may serve as an illustration.

On the other hand, since we have shown that the distinction between $\langle N \rangle_k$ and $\bar{N}$ can be traced to the width of the distribution $P(N)$, a fit to the resulting deviations may be employed to characterize the size distribution function. This may serve as a new means of probing nanocluster and nanodroplet nucleation in beam expansion sources.

*Acknowledgments.* We would like to thank Dr. Michal Fárník for valuable comments. This work was supported by the U.S. National Science Foundation under Grant No. CHE-2153255.

*Author declarations.*

Conflict of interest. The authors have no conflicts to disclose.

Data availability. The data that support the findings of this study are available from the corresponding author upon reasonable request.

Author contributions. **Marie Riddle**: Formal analysis, methodology, software, writing - review and editing; **Angel I. Pena Dominguez**: Formal analysis, methodology, software, writing - review and editing; **Benjamin S. Kamerin**: Investigation, conceptualization, writing - review and editing; **Vitaly V. Kresin**: Conceptualization, formal analysis, writing – original draft, funding acquisition.

**Appendix**

**1.** Fig. 3 exhibits how a change in the pick-up vapor pressure shifts the subpopulation of helium nanodroplets that pick up only a single dopant. The data set obtained with a higher dopant vapor pressure displays a stronger deflection, reflecting a decrease in the average size $\langle N \rangle_1$ of the host nanodroplets (cf. Fig. 1).

**2.** The log-normal distribution of cluster sizes is given by[24,25]

$$P(N) = \frac{1}{N\delta\sqrt{2\pi}} \exp\left[-\frac{1}{2}\left(\frac{\ln N - \mu}{\delta}\right)^2\right],$$

with $\mu = \ln\left(\bar{N}/\sqrt{1+r^2}\right)$, $\delta = \sqrt{\ln(1+r^2)}$. Here $r = s/\bar{N}$ is the ratio of the standard deviation of $N$ to the mean.

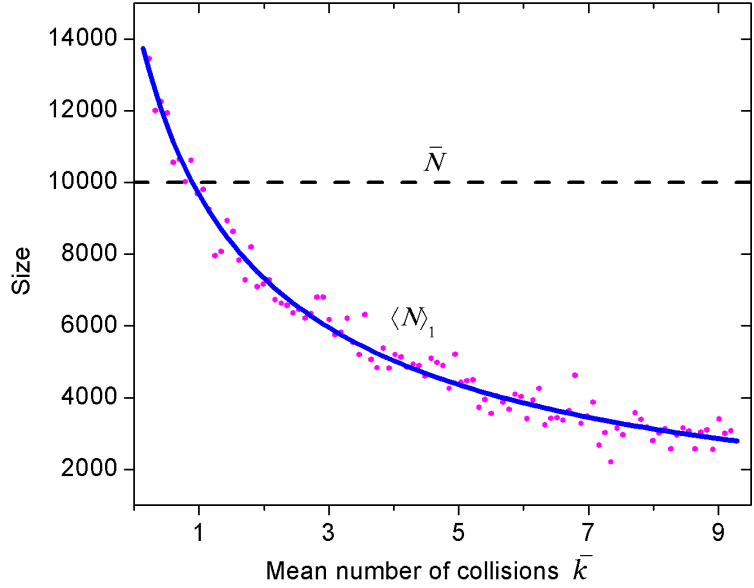

**Figure 1**. The average size of nanodroplets carrying one dopant after passing the pick-up region (solid line: Eq. (3), dots: Monte Carlo simulation), compared with the mean nanodroplet size in the original beam (dashed line). The horizontal axis marks the mean number of nanodroplet collisions in the pick-up region for a nanodroplet of size $\bar{N}$.

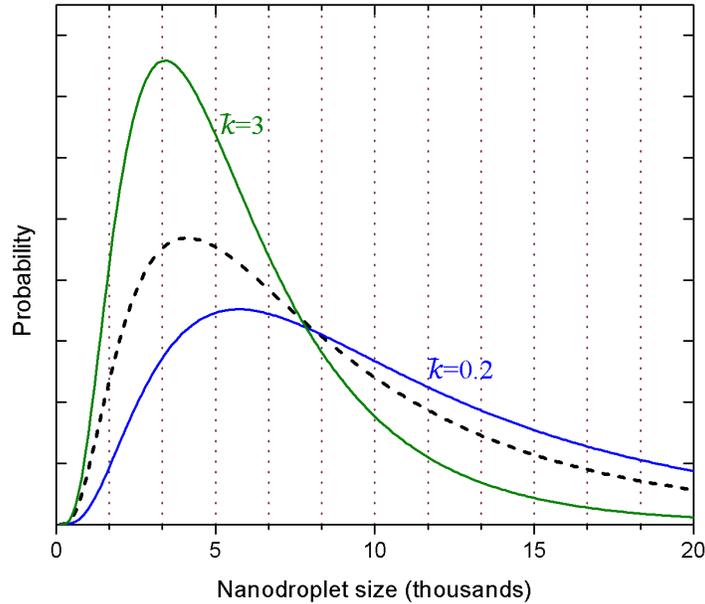

**Figure 2**. Dashed line: original log-normal nanodroplet size distribution (same as employed for Fig. 1). Solid lines: populations of nanodroplets carrying one dopant after traversing the pick-up region in low ($\bar{k}=0.2$) and high ($\bar{k}=3$) vapor density regimes.

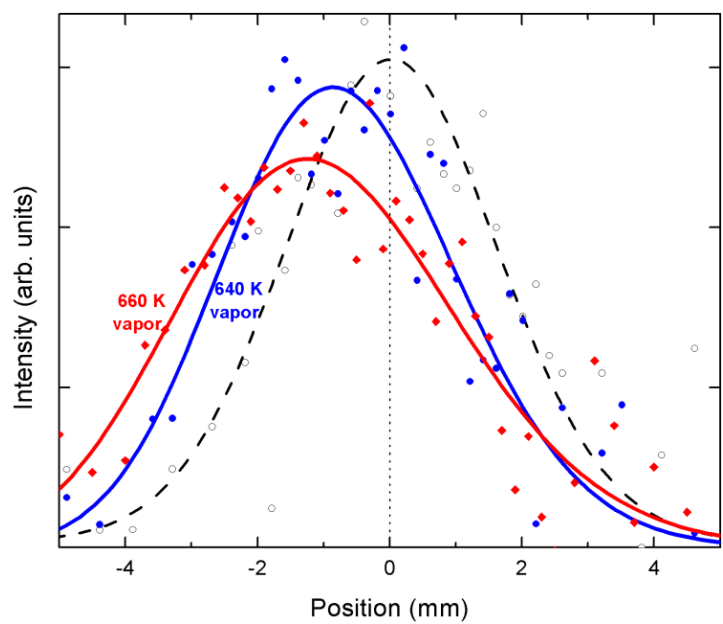

**Figure 3**. Stern-Gerlach deflection profiles[21] of helium nanodroplets doped with a single (i.e., *k*=1) high-spin molecule $FeCl_2$ obtained at two temperatures of the vapor pickup cell. The curves are fits to the data (symbols), the dashed curve is the zero-field profile centered on the beam axis (dotted line).